# Disruption of the Orion Molecular Core 1 by the stellar wind of the massive star $\theta^1$ Ori C

C. Pabst[1], R. Higgins[2], J.R. Goicoechea[3], D. Teyssier[4], O. Berne[5], E. Chambers[6], M. Wolfire[7], S. Suri[2], R. Guesten[8], J. Stutzki[2], U.U. Graf[2], C. Risacher[8,9], A.G.G.M. Tielens[1]

[1]Leiden Observatory, Leiden University, Niels Bohrweg 2, 2333CA Leiden, The Netherlands
[2]I. Physikalisches Institut der Universität zu Köln, Zülpicher Strasse 77, 50937 Koln, Germany
[3]Instituto de Física Fundamental (CSIC), Calle Serrano 121, 28006 Madrid, Spain
[4]Telespazio Vega UK Ltd for ESA/ESAC, Camino bajo del Castillo, s/n, Urbanizacion Villafranca del Castillo, Villanueva de la Cañada, E-28692 Madrid, Spain
[5]IRAP, Université de Toulouse, CNRS, CNES, Université Paul Sabatier, (Toulouse), France
[6]USRA/SOFIA, NASA Ames Research Center, Mail Stop 232-12, BuildingN232, PO Box 1, Moffett Field, CA94035-0001, USA
[7]Department of Astronomy, University of Maryland, College Park, MD 20742, USA
[8]Max-Planck-Institut für Radioastronomie, Auf dem Hugel 69, 53121 Bonn, Germany
[9]IRAM, 300 rue de la Piscine, 38406, St. Martin d'Hères, France

Mechanical and radiative energy input by massive stars stir up the environment, heat the gas, produce cloud & intercloud phases in the interstellar medium and disrupt molecular clouds, the birthsites of new stars[1,2]. Ionization by UV photons, stellar wind action and supernova explosions control molecular clouds lifetimes[3,4,5,6,7]. Theoretical studies predict that momentum injection by radiation dominates by far over momentum injected by a stellar wind[8], but this has hitherto been difficult to assess observationally. Velocity-resolved large-scale images in the fine structure line of ionized carbon ([CII]) provide an observational diagnostic of the radiative energetics and the dynamics of the ISM in the immediate vicinity of massive stars. Here, we present the [CII] 1.9 THz (158 μm) study of ~1 square degree region (~7pc in diameter) at a resolution of 16" (0.03pc) of the nearest region of massive star formation, Orion. The results reveal that the stellar wind originating from the star, $\theta^1$ Ori C, has created a ~2pc sized bubble by sweeping up a 2600 $M_\odot$ shell expanding at 13 km/s. This shows that the stellar wind mechanical energy is coupled very efficiently to the molecular core and its action dominates over photo-ionization/evaporation or future supernova explosions.

We have surveyed one square degree of the Orion Molecular cloud, centered on the Trapezium cluster and the Orion Molecular Core 1 (OMC-1) behind it, in the 1.9 THz (158µm) [CII] fine-structure line with the 14 pixel upGREAT heterodyne high-spectral resolution spectrometer[9] on board of the Stratospheric Observatory For Infrared Astronomy (SOFIA) (see method). Figure 1 compares the [CII] integrated intensity map with the mid-IR and far-IR maps due to UV-pumped fluorescence by polycyclic aromatic hydrocarbon (PAHs) molecules and thermal dust continuum emission, respectively. Each map clearly shows the direct interaction of the Trapezium cluster with the dense molecular core (center), the large, wind-blown bubble, associated with the Orion Veil (South), and the bubble created by the B stars illuminating the reflection nebulae, NGC 1973, 1975, and 1977 (North). Here, we focus on the prominent Veil bubble associated with the stellar wind from $\theta^1$ Ori C. This shell consists of neutral atomic (H) gas and is very prominent in the [CII] map but there is no detectable counterpart in carbon monoxide, $H_2$, or other molecular tracers as the shell is too tenuous for these species to persist; e.g., $H_2$/H fraction <$2 \times 10^{-4}$ and $C/C^+ = 10^{-4}$ [10,11]. Likewise, the complex pattern of absorption and emission features and the presence of multiple (foreground) components preclude recognition of the large scale structure of the shell in 21cm HI studies[12]. X-ray observations[13] have shown that this bubble is filled with tenuous (~1$cm^{-3}$) hot ($2 \times 10^6$ K) gas created by the strong stellar wind (mechanical luminosity, $L_w = 8 \times 10^{35}$ erg/s[14,15]) from the most massive star in the region, $\theta^1$ Ori C (see Extended Data Figure 5).

While each IR image (Fig. 1) traces the Veil morphology, only [CII] probes the kinematics through the Doppler effect. The high spectral resolution of upGREAT allows a detailed investigation of the gas dynamics, revealing the kinematic signature of an expanding half shell as, with increasing velocity, the shell displaces further away from the center (Extended Data Figure 6). The kinematic data show good agreement of the observed velocity structure with a simple model of a half-shell expanding at 13 km/s towards us while expansion into OMC-1 is stopped by its high density (n=$10^4$–$10^5$ $cm^{-3}$) (Fig. 2; Extended Data Figure 7). The small velocity difference (~1 km/s towards us) between $C^+$ and CO emission from the OMC-1 gas represents a slow photo-evaporative flow of atomic gas (H, $C^+$) into the bubble, where H is then ionized by EUV (E>13.6 eV) photons from $\theta^1$ Ori C before flowing into the cavity at ~17 km/s[16]. We have determined the mass of gas in the Veil to be between 1700 and 3400 $M_\odot$ with a most likely value of 2600 $M_\odot$ from an analysis of the far-IR dust emission (see Method). Analysis of the weak [$^{13}$CII] hyperfine line component apparent after averaging over the shell results in a very similar value. This mass estimate is about twice the mass derived from the HI column densities observed along pencil-beams towards the Trapezium stars, which likely reflects known fluctuations in the shell thickness in these directions. The derived Veil mass is comparable to the mass of gas in OMC-1 (~3000 $M_\odot$)[17] and the mass of the newly formed stellar cluster (~1800 $M_\odot$)[18], and much exceeds the mass of ionized gas (2 $M_\odot$ in the dense Huijgens region and 20 $M_\odot$ in total)[19] and the mass of the X-ray emitting hot plasma (0.07 $M_\odot$[13]; Extended Data Table 1). A schematic view of the region is shown in Figure 3.

Adopting a homogeneous cloud, and a size of 2 pc, the mass of swept-up material corresponds to an initial $H_2$ density of $1.4 \times 10^3$ $cm^{-3}$. The radius of the shell is given by $R_s(t) = (125/154\pi)^{1/5} (L_w/\rho_o)^{1/5} t^{3/5}$ with $L_w$ the wind mechanical luminosity, $\rho_o$ the initial density and $t$ the time[20] and we derive an age of 0.2 Myr. This age is in the range of earlier (uncertain) estimates of $3 \times 10^4$–$10^6$ yr based on the expansion of the

HII region, the ages of proplyd stars and the Orion Nebula cluster stars[18,21], but exceeds the typical dynamical lifetime expected for trapezium-type, multiple systems, $1-5\times10^4$ yr[22]. The derived lifetime of the bubble with the mass loss rate of $\theta^1$ Ori C implies a total stellar mass injected of 0.08 $M_\odot$ close to the mass of the hot plasma estimated from the X-ray observations[13].

Mass, energy, and luminosity of the shell are compared to those of other relevant components in Table 1 (Extended data). The total kinetic energy of the expanding half shell is $\sim4\times10^{48}$ erg, comparable to the total mechanical energy delivered by the wind over the age of the bubble ($5\times10^{48}$ erg). The total kinetic energy in the ionized gas, $6\times10^{46}$ erg, is much less than the kinetic energy of the wind bubble. The observed X-ray luminosity over the age of the bubble is only $3\times10^{44}$ erg and the hot gas expands adiabatically. Assuming that the hot gas fills the cavity, its thermal energy is only $\sim10^{47}$ erg[13]. Theory predicts that 5/11 of the mechanical energy of the stellar wind will go into heating the hot gas and 6/11 goes into work done on the environment[20] and this discrepancy may indicate that the bubble is leaking hot gas into the surrounding Orion-Eridanus superbubble[13]. However, while detailed inspection of the shell reveals that it is quite thin at places, the position-velocity diagrams provide no clear kinematic signature of rupture, which would show up as a "local" rapid variation in velocity. The radiative luminosity of $\theta^1$ Ori C over 200,000 yr is $6\times10^{51}$ erg. The kinetic energy of the ionized gas corresponds then to a coupling efficiency of $10^{-5}$ for the coupling of this energy to radial momentum. This efficiency is in good agreement with theoretical studies on this coupling for clouds phases of the ISM[8]. We also note that the radiation pressure ($4\times10^4$ cm$^{-3}$ K) is well below the thermal pressure of the hot gas ($10^6$ cm$^{-3}$ K) and that the Ori Id cluster is very young (<1 Myr) and massive stars have not had the time to evolve to the SN stage and neither of these processes will have played a role in the creation of the Veil bubble.

The slow shock propagating into the environment during the expansion will heat swept up molecular gas to a $\sim3000$ K and this internal energy is quickly radiated away in high J CO and pure rotational $H_2$ lines. Such low velocity shocks do not dissociate or ionize molecular gas or emit in the [CII] line[23]. The observed total luminosity radiated by dust in the swept up shell is $\sim6\times10^4$ $L_\odot$, resulting in $\sim1.5\times10^{51}$ erg over the Veil expansion time. The luminosity of $\theta^1$ Ori C provides this energy as stellar photons travel unimpeded through the evacuated cavity, illuminating the inner bubble boundary. This excavation turns the [CII] line and the far- and mid-IR tracers into good tracers of the shell (Fig. 1). Observations have shown that gas illuminated by strong radiation fields typically emit between 0.5 and 2% of the stellar photon energy in the [CII] 1.9 THz gas cooling line[24,25]. Stellar FUV radiation field is coupled to the gas through the photo-electric effect on PAH molecules and very small grains[26]. The observed [CII] emission ($L_{C+}=200$ $L_\odot$) for the Orion Veil translates into a photo-electric efficiency of 0.3%, well in line with these studies. UV photon energy that does not go into ionization, emerges as the PAH emission features that so prominently outline the shell (Fig 1).

The derived velocity of the shell (13 km/s) exceeds the escape velocity of OMC-1 ($\sim2$ km/s) and of the Orion Molecular Cloud A ($\sim8$ km/s). Eventually, the wind bubble will break open and vent the hot gas as well as the ionized gas into the surrounding, tenuous Orion-Eridanus superbubble (see method). The coasting, neutral shell will then dissolve into the hot plasma. Supernovae typically go off every $\sim1$ Myr in the

Orion OB1a and 1b sub-associations and sweep up all the "loose" material deposited in the superbubble by bursting bubbles such as the Veil and transport this material to the wall of the superbubble[27]. In essence, supernova mechanical energy (~$10^{51}$ erg) will go into rejuvenation of the hot gas in the superbubble and transportation of swept up gas towards its walls and very little will couple to the Orion Molecular Clouds A and B. Barnard's loop may be the latest "episode" in this process[27]. Considering $\theta^1$ Ori C, estimates for its proper motion with respect to the molecular cloud range from ~5 to 15 km/s, with the latest evidence pointing towards the lower value[28] and $\theta^1$ Ori C will move ~25 pc away from the cloud before it explodes as a supernova. Hence, like the Orion OB 1a and 1b stars, as a supernova, $\theta^1$ Ori C will not affect the evolution of its "birth" core. Recent, three-dimensional hydrodynamic simulations also reveal, from a theoretical side, that stellar winds are key to the regulation of star formation through their action on molecular clouds[29]. Here, we have analyzed one specific case of the interaction of a wind from a massive star with its environment and the general nature of this conclusion still needs to be assessed. [CII] 1.9 THz observations with SOFIA provide a premier tool for such studies.

Galaxy formation and evolution result from the combined effects of a complex set of physical processes that affect the baryons in a ΛCDM cosmology dominated by dark matter. In particular, stellar feedback controls the evolution of galaxies[30]. Stellar winds from O-type massive stars are very effective in disrupting molecular cores and star formation and, as stellar wind energy input is dominated by the most massive stars in a cluster while SN energy input is dominated by the more numerous B-type stars, this has a direct impact on cosmological simulations. As our study shows, relevant stellar feedback processes act on much smaller scales (0.2–2pc) than hydrodynamic studies of ISM evolution (>2pc) or cosmological simulations (>50 pc) resolve[2,4,5,6,7]. [CII] 1.9 THz studies on the dynamic interaction of massive stars through stellar winds with nearby molecular clouds can provide key validation for theoretical studies.

**Acknowledgements:** The authors gratefully acknowledge the dedication and hard work of the USRA and NASA staff of the Armstrong Flight Research Center in Palmdale and of the Ames Research Center in Mountain View as well as the DSI during the upGREAT Square Degree Survey of Orion. Research on the Interstellar Medium at Leiden observatory is supported through a Spinoza award. We thank the ERC and the Spanish MCIU for funding support under grants ERC-2013-Syg-610256-NANOCOSMOS and AYA2017-85111-P, respectively.

**Authors contributions:** J.R.G. D.T., O.B., M.W., and A.G.G.M.T conceived the Orion Square Degree Survey and wrote the proposal for SOFIA. R.H., D.T., E.C., and J.R.G. optimized the observing strategy for this large survey. R.G., J.S., U.U.G., R.H., C.R., and C.P. carried out the observations. E.C. was responsible for the link to the SSC. R.H., assisted by C.P., was responsible for the data reduction. C.P. was responsible for the analysis and interpretation of the [CII] as well as the Herschel data. S.S. compared the [CII] data with molecular observations. A.G.G.M.T. provided overall guidance and wrote the paper with contributions from all coauthors.

**Author Information:** Reprints and permissions information is available at www.nature.com/reprints. The authors declare no competing financial interests. Readers are welcome to comment on the online version of the paper. Publisher's note: Springer Nature remains neutral with regard to jurisdictional claims in published maps and institutional affiliations. Correspondence and requests for materials should be addressed to A.G.M.T. (tielens at strw.leidenuniv.nl).

**Figure legends:**

**Figure 1**: **Three infrared views of the Orion region of massive star formation.** Each of these images reveals very similar morphology but in different tracers of the dust and gas in the molecular cloud. **a)** The dust continuum view as observed by the Herschel Space Observatory in far-infrared (blue; PACS) and sub-millimeter (red; SPIRE) emission, measuring the conversion of FUV radiation from massive stars to dust emission in the photodissociation region (PDR). **b)** The integrated [CII] 1.9 THz (158 μm) emission as observed by the upGREAT instrument on board of the Stratospheric Observatory For Infrared Astronomy, tracing cooling and kinematics of PDR gas. **c)** The 8 μm PAH emission observed by the IRAC instrument on board of the Spitzer Space Telescope, outlining the FUV illuminated PDR surfaces. This comparison does not do full justice to the richness of the [CII] data as there are actually ~2,200,000 spectra that turn this 2D image into a 3D view and allows a detailed study of the kinematics of the gas.

**Figure 2: Position–velocity diagrams of the [CII] emission along selected cuts across the Veil. a)** The Veil bubble in the integrated intensity of the [CII] 1.9 THz emission. The red lines delineate the region over which the spectra were collapsed to produce the East-West cross cut shown in the panels (b) and (c). The 0,0 position corresponds to the position of $\theta^1$ Ori C (RA(2000)=05 35 16.46 Dec(2000)=-05 23 22.8) and is indicated by a yellow star. The orange star indicates the position of the unrelated star, $\theta^2$ Ori A (RA(2000)=05 35 22.90 Dec(2000)=-05 24 57.82). **b)** Position-velocity (p-v) diagram of the [CII] emission in the East-West crosscut indicated in panel (a). Other horizontal and vertical crosscuts show similar p-v diagrams (Extended Data figure 7). **c)** A simple model of a spherical half shell expanding at a constant velocity of 13 km/s is fitted to the observed p-v data. All of the observed p-v diagrams are in good agreement with this simple model of a half sphere expanding at a velocity of 13 km/s.

**Figure 3: Sketch of the structure of the Orion stellar wind bubble.** The stellar wind from the massive star, $\theta^1$ Ori C, is shocked by the reverse shock, creating a hot (~$2 \times 10^6$ K), tenuous (~1 $cm^{-3}$), X-ray emitting plasma. Adiabatic expansion of this hot gas has swept up the surrounding gas into a slowly expanding (13 km/s) 4pc diameter half-shell. The dense Orion molecular cloud core, OMC-1, behind the Trapezium cluster stops expansion of the bubble in this direction. Photons with energies above 13.6 eV can ionize H and create the dense ionized gas layer, the Huygens region, behind the Trapezium cluster, which dominates optical images of the region. This ionized gas expands into the bubble at ~17 km/s with respect to the background molecular gas. The largely empty interior of the hot gas bubble allows UV photons from $\theta^1$ Ori C to travel unimpeded until they interact with the gas and dust in the shell. These photons will heat the dust, leading to bright far-IR continuum emission, and excite large polycyclic aromatic hydrocarbon molecules, producing the 8 μm fluorescence emission (see figure 1). Far-UV photons will ionize carbon, and heat the largely neutral gas to ~200 K. This gas cools through the [CII] 1.9 THz line. On a much larger scale (~350 pc), the Orion molecular cloud and the Veil bubble are embedded in the Orion-Eridanus superbubble (not to scale).

**Extended data figure Legends:**

**Extended Data Figure 1: Outline of the region mapped in the [CII] 1.9 THz line with upGREAT on SOFIA.** The 78 tiles indicated were used to construct the final map. Background image is a 70 micron Herschel/PACS dust emission. The yellow contours correspond to an approximated far-UV radiation field of $G_0$=50 (in Habing units). The color of each tile indicate its corresponding OFF position, blue tiles use the COFF-SE1 position, red tiles COFF-OFF1 and green tiles COFF-C. Each square tile has a side length of 435.6 arcseconds. The black box at the center indicates the region mapped by the single pixel Herschel/HIFI instrument in a time of 9 hours[54]. The total observing time for the SOFIA/upGREAT map was 42 hours.

**Extended Data Figure 2: Sample [CII] 1.9THz spectra in our data cube. a)** Spectrum obtained at the map center (RA: 5h35m17s Dec:-5dm22d16.9). **b)** Averaged spectrum over the entire map.

**Extended Data Figure 3**: **Schematic overview of the large scale (~350 pc) structure of Orion.** We mark with green stars the locations of the massive stars spanning up the Orion constellation (shoulders and knees, but the belt is indicated by a single star; M42 is at the tip of the sword), in blue outline the two giant molecular clouds A & B, in green filled trace the prominent HII regions including M42 powered by the trapezium cluster, in red Barnard's loop, which is very prominent in Hα. The bubble surrounding λ Ori is also indicated (red=ionized gas, blue swept up molecular shell, and the boundaries of the superbubble (yellow). The locations of the Orion OB subassociations are marked in green.

**Extended Data Figure 4: Overview of the star forming region in Orion.** The approximate boundaries of the Orion OB associations Ib, & Ic are indicated by dashed ellipses. The Orion Id association is directly associated with the molecular cloud behind the Orion Nebula, M42. The reddish glow is due to the Hα line originating from recombinations in the ionized gas of Barnard's loop. The belt stars and the knees are obvious. The size of the image is approximately 10 degrees on the sky.

**Extended Data Figure 5: Composite infrared and X-ray views of the Orion region of massive star formation.** The [CII] integrated intensity map is shown on a color scale. The X-ray emission is outlined by a green contour. Likely, the hot gas fills the bubble on its entirety but absorption by the Veil preferentially extinguishes the left side. The position of the star, $\theta^1$ Ori C (RA(2000)=05 35 16.46 Dec(2000)=-05 23 22.8) is indicated by a blue star. Credit: X-ray data XMM-Newton.

**Extended Data Figure 6: Composite figure showing the [CII] emission in different velocity channels.** Note that with increasing $v_{LSR}$, the shell is displaced outward, away from the bubble center. This is the kinematic signature of an expanding half shell. Each color outlines the emission boundaries of 1 km/s wide channels from $v_{LSR}$ = 0 to 7 km/s. The 0,0 position corresponds to the position of the exciting star, $\theta^1$ Ori C (RA(2000)=05 35 16.46 Dec(2000)=-05 23 22.8) and is marked by a magenta star. Note that in the velocity range, 4 to 7 km/s, [CII] emission associated with the OMC-4 core starts to fill in the bubble interior. OMC-4 is a starforming core near the front of the background molecular cloud and is not part of the Veil bubble.

**Extended Data Figure 7: Four exemplary position–velocity diagrams (pv) of the [CII] emission along selected cuts across the Veil.** Each pv diagram exhibits a clear arc structure extending over ~2500", corresponding to the expanding Veil shell (Pabst et al 2018, in preparation). The left two panels are cuts along the horizontal axis. The right two panels show cuts along the vertical axis.

**Extended Data Figure 8: Far infrared dust emission in Orion**. **a)** Optical depth map of the dust emission at 160 μm, $\tau_{160}$, tracing the mass of the shell. The two big circles indicate the extent of the shell used to determine the mass of the limb brightened shell. The small circle, inscribed `OMC1' circumscribes the Huijgens region associated with the Trapezium stars. We have estimated the mass that is enclosed between these circles, excluding the OMC1/Huijgens region. **b)** Spectral energy distribution (SED) of the dust emission observed for different positions in Orion. These SEDs are analyzed to determine the dust and gas mass. Data points and curves represent observed SEDs and model fits for $\beta$ = 2. The legend provides resulting dust temperature $T_d$ and dust optical depth, $\tau_{160}$, values. These SED fits have been analyzed for each spatial point and the resulting optical depth values have been used to construct the map shown in panel a.

**Extended Data Figure 9: Average spectra from the shell.** These spectra are dominated by the [CII] line from the main isotope and show the weak hyperfine component of $^{13}C^+$ near $v_{LSR}$ ~ 20 km/s. This line can be used to estimate the optical depth of the main isotope line and then the mass of emitting gas. The red spectrum corresponds to the area between the two large circles in Fig. 1 but excluding region in the circle encompassing the OMC1/Huygens region. The inset of the spectrum shows the zoom in on the (faint) [$^{13}$CII] line in the average shell spectrum. The blue spectrum is an average over the bright parts in the eastern shell, in the declination range $\delta$ = -5°35' to -5°45'.

**Extended Data Table 1: Masses, energetics and luminosities in Orion**

**Figures**

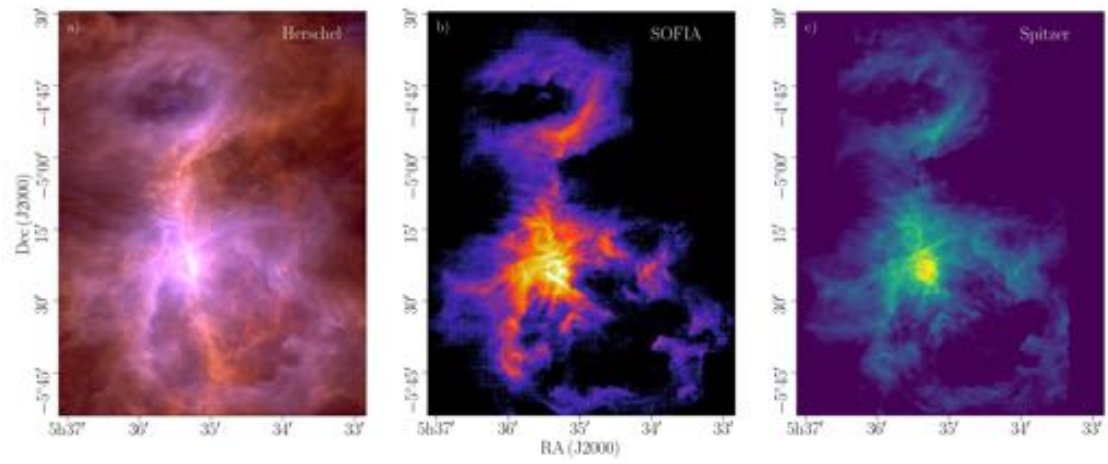

Figure 1

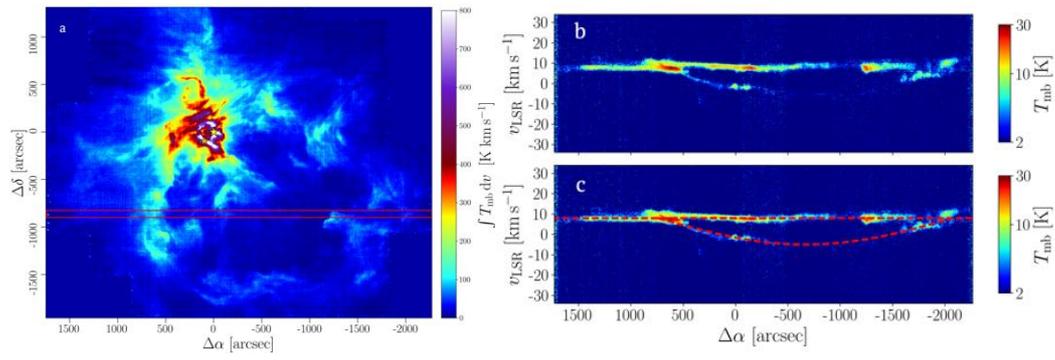

Figure 2:

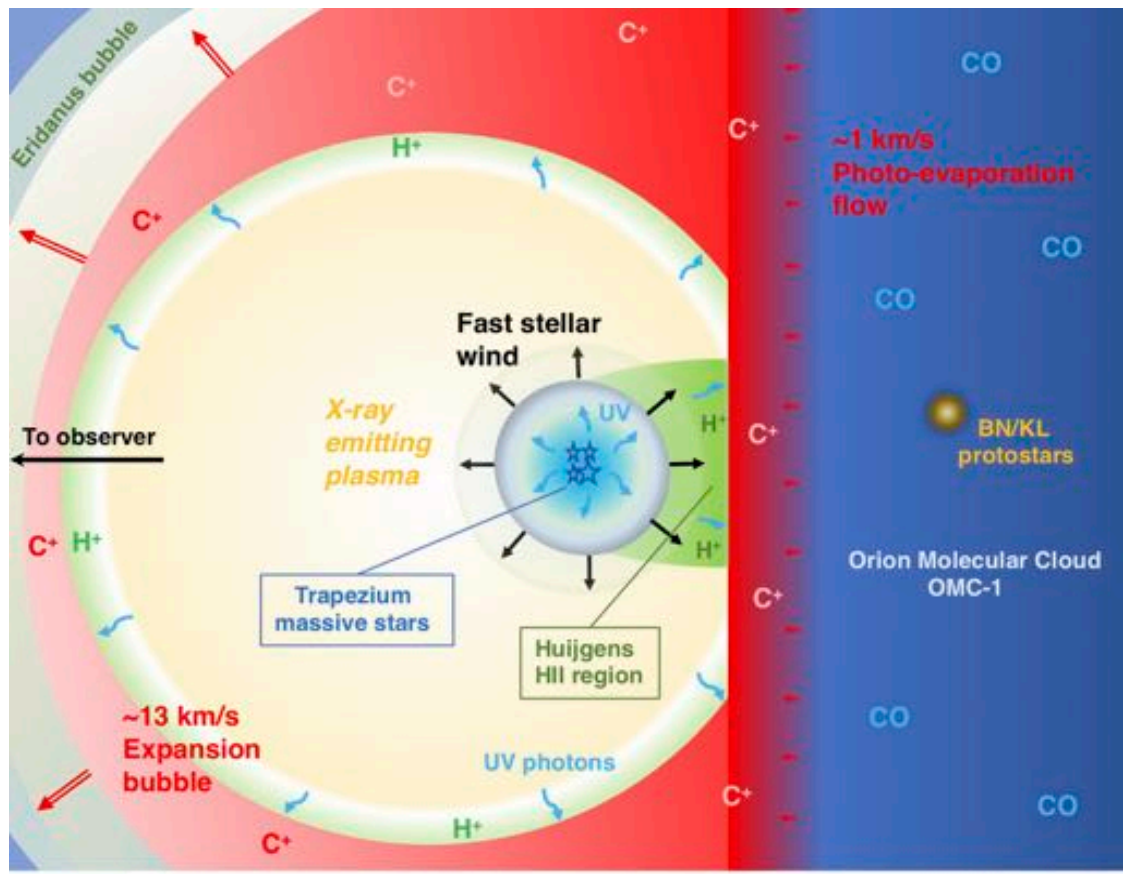

Figure 3

# Methods

**SOFIA observations**

The data presented in this article were taken using the upGREAT (German Receiver for Astronomy at Terahertz Frequencies) heterodyne receiver flying abroad the SOFIA (Stratospheric Observatory for Infrared Astronomy) observatory[10]. SOFIA is an 80/20 joint project between NASA (National Aeronautics and Space Administration) and the DLR (German Aerospace Centre). SOFIA is a modified Boeing 747-SP aircraft with a 2.5 diameter telescope mounted in the fuselage aft of the wing[31]. Flying at up to 43,000 feet (13 km) altitude provides access to frequencies typically absorbed by the atmosphere.

The upGREAT receiver is heterodyne array receiver with 21 pixels. 2 hexagonal arrays each with 7 pixels are found in the low frequency array (LFA), covering a frequency range from 1.81 THz to 2.08 THz. The other 7 pixels are found in the high frequency array (HFA) in a similar hexagonal pattern to the LFA. The HFA is tuned primarily to the atomic oxygen transition at 4.744 THz. A heterodyne receiver has a spectral resolving power up to $\nu/\Delta\nu = 10^7$ by mixing a locally generated monotone signal close in frequency to the astronomical signal of interest with the broadband sky signal. The beat tone between the 2 signals contains the astronomical signal but at microwave frequencies that can be amplified and sampled using microwave (GHz) electronic components. The original data at a resolution of ~0.04 km/s was rebinned to 0.2 km/s to increase signal-to-noise ratio.

The data was taken using a special array on-the-fly (OTF) mapping mode. This differs from a classical OTF mode where a single pixel is traced through the map dimensions and a fully sampled map in generated. In the array-OTF observing mode, we use the unique hexagonal array geometry to generate a fully sampled map. With this approach each receiver pixel doesn't cover every map point, however we can map larger regions in the same time as a classical OTF approach. However we lose some signal to noise and pixel redundancy. This mode is described in more detail in ref 10.

The full map region was broken in 78 square tiles of length 435.6 arcseconds (Extended Data Figure 1). Each tile took 22 minutes to complete. A tile is made up of 84 scan lines separated by 5.2 arcseconds. Each tile is covered twice, once in the X and once in the Y direction. Each OTF scan line is made up of 84 dumps of 0.3 seconds. This returns noise RMS for a spectral resolution of 0.3km/s of 1.14 K ($T_{mb}$) per map pixel.

The raw data is recorded by a digital spectrometer and comes in the form of integers counts per spectrometer channel. These values are converted to antenna temperature using an internal hot and cold reference measurements which establish a scale for the sky measurements. The observation of detected off positions, free from CII emission, is required to remove instrument artifacts from the data.

The next step in the calibration process is to establish the atmospheric transmission and apply this to the astronomical signal. While SOFIA flies above most of the atmosphere. There are some atmospheric features that need to be considered in the final calibration. The atmospheric emission is determined by fitting an atmospheric model to the OFF minus internal hot data. The process of atmospheric determination is described in the detail in ref 32.

Once the atmospheric transmission is applied to each spectrum, a channel map can be generated. In total 2.2 million spectra were taken. This dataset is converted to a map[33] by defining a map grid and each map pixel is generated by the distance weighted sum of all spectra within a given distance of the pixel center. This begins an iterative process where map artifacts are identified and then a correction is applied to the individual spectra and the map is regenerated.

While the data quality from the upGREAT instrument was exceptional, with 90% of spectra requiring no post-processing, some spectra did require some post processing. Nominally, one could drop problematic spectra and still have enough spectra for a completely sampled map however with the array OTF mapping mode this may not be the case. In order to recover problematic spectra we developed a spline correction method. The classic approach in heterodyne data processing would be to use a polynomial to remove spectra "baseline" artifacts. However this can be problematic and is difficult to constrain. We have adapted an approach pioneered by the Herschel/HIFI instrument[34,35], which used a catalog of splines generated from data with no astronomical signal. These splines can then be scaled to astronomical data and produces a more effective baseline removal. Representative spectra are shown in Extended Data Figure 2.

**Orion**. Orion is the nearest site of massive star formation and has long been used as a laboratory for the study of the interaction of massive stars with molecular clouds[36,37]. The region contains two molecular clouds, Orion Molecular Cloud A and B. The Orion Molecular Core 1 is one of four cores that have condensed out in the so-called integral-shaped filament that forms the densest part in the Orion Molecular Cloud A. The distance to the Orion Molecular Cloud A is known to vary by about 30pc on a scale of 50 pc[38], but on the scale of the Orion Nebula Cluster, the HII region, M42, and the Veil region, the distance is well determined at 414±7 pc[39,40]. This small uncertainty in the distance of OMC1 does not affect the main results of the paper regarding mass or kinetic energy.

The Orion OB associations represent the effects of ongoing formation of (massive) stars in this region and their interaction with the environment over some 10 Myr[41]. The oldest sub-associations, Orion OB1a and OB1b, consist of the stars in the Orion Belt and just North of it (see Extended Data figures 3 & 4 for an overview of the region). These subgroups have produced several supernovae that have swept up their environment, creating the 350 pc diameter Orion-Eridanus superbubble. The Orion OB1c subgroup in the sword is somewhat younger (5-8 Myr), while the youngest (<1Myr) stellar subgroup, Orion OB1d, represents still active massive star formation associated with the prominent HII regions, M42 (the Orion Nebula), M43, and NGC 1977. Part of this subgroup is still embedded in the Orion A molecular core, OMC-1.

The Trapezium cluster, $\theta^1$ Ori, is located on the front side of OMC-1. Each of the four stars making up the trapezium cluster is a multiple system in itself. $\theta^1$ Ori C is a binary where the primary has a mass of 34 $M_\odot$ and the companion is only 12 $M_\odot$[42]. The primary's spectral type is O7Vp with an effective temperature of 39,000K. $\theta^1$ Ori A and $\theta^1$ Ori D are both much lower mass (14 and 16 $M_\odot$) with spectral types B0.5V. Both are also binary systems with a lower mass companion[43]. $\theta^1$ Ori B is only 7 $M_\odot$[44]. The radiative energy input in the region is dominated by far by the most massive star, $\theta^1$ Ori C as the other stars contribute less than 20% of the luminosity of the region. $\theta^1$ Ori C also dominates by far (>90%) the ionizing radiation of the cluster.

θ¹ Ori C also has a strong stellar wind with a mass loss rate of 4x10⁻⁷ M$_\odot$/yr and a terminal velocity of 2500 km/s corresponding to a mechanical luminosity, $L_w$=8x10³⁵ erg/s[15,16]. Older estimates give a mechanical luminosity, $L_w$=7x10³⁵ erg/s[45], but this small difference has no influence on our discussion. B0.5 stars have very weak stellar winds and the mechanical energy input by the other trapezium stars is negligible. The wind from θ¹ Ori C has blown a bubble filled with hot, tenuous gas (density, n~ 1 cm⁻³, and temperature, T~2x10⁶ K)[14]. This hot gas dominates the diffuse emission at X-ray wavelengths (Extended Data Figure 5). This bubble is well outlined in the mid-IR image obtained in the polycyclic aromatic hydrocarbon (PAH) emission bands and in the far-IR image of the dust thermal emission (figure 1). These images trace the interaction of far-UV (<13.6eV) radiation from θ¹ Ori C with these species in the neutral photodissociation region (PDR) separating the hot bubble gas from the cold molecular cloud material.

**Kinematics of the gas**

Analysis of the individual velocity channel maps reveals the kinematic signature of an expanding half shell as with increasing velocity, the shell displaces further away from the center (Extended Data Figure 6). We estimate the expansion velocity of the expanding shell of the Orion Nebula from position-velocity (pv) diagrams. Across the range of the velocity-resolved [CII] map shown in Figure 2, we build pv diagrams by averaging spectra over cuts that are each 45.5" wide, both along the horizontal and vertical directions (Pabst et al, 2018, in prep.). Figure 2 shows a representative example of one of these pv diagrams. We illustrate the results further with four pv diagrams, shown in Extended Data figure 7. Details will be discussed in Pabst et al, 2018, in prep. The expanding shell shows clearly as an arc structure that is visible in most of the cuts. From single pv diagrams that most significantly exhibit this arc we estimate the expansion velocity by calculating the centroid velocity within the velocity ranges 5 to 15 km/s for the [CII] background velocity (at the surface of the molecular cloud) and -10 to 5 km/s for the expanding gas, respectively: the difference is the expansion velocity with respect to the background molecular cloud. The entirety of [CII] pv diagrams are well described by an expanding-shell model with one central origin and an expansion velocity of 13±1 km/s (Pabst et al 2018, in prep.). Corresponding ¹²CO(2-1) and ¹³CO(2-1) pv diagrams[23] along the same spatial cut do not show any sign of an expanding shell. This is in line with the low concentration of molecules in the Veil derived from UV absorption line studies towards the trapezium stars[11]. The pv diagrams reveal that molecular gas is shifted towards slightly higher velocity as compared to the [CII] line (11km/s versus 9 km/s).

**Mass estimates of the Veil**

We have estimated the gas mass in the expanding shell from fits to the far infrared observations of the dust emission using standard dust-to-gas conversion factors based on the models of reference 27. We have used Herschel far-infrared photometric images in the PACS bands at 70, 100, and 160 μm, and the SPIRE 250, 350, and 500 μm bands. We have convolved all images to the spatial resolution of the SPIRE 500 μm image (36"). We have then fitted the spectral energy distribution (SED) per pixel, using a modified blackbody distribution,

$$I(\lambda) = B(\lambda, T)(1 - e^{-\tau(\lambda)})$$

with

$$\tau(\lambda) = \tau_{160} \left(\frac{\lambda}{160\ \mu m}\right)^{-\beta}$$

where $B(\lambda, T)$ is the Planck black body spectrum at temperature, $T$, and wavelength, $\lambda$, and $\tau_{160}$ the optical depth at 160 μm. The dust optical depth varies with wavelength and this is parametrized by the grain emissivity index $\beta$. Typically, $\beta$ is in the range 1 to 2. Here, we have chosen $\beta = 2$, in accordance with the standard dust models for $R_V$ = 5.5, appropriate for the Orion molecular cloud (for details, see Pabst et al 2018, in prep.). Representative fits to the SEDs and the resulting map of the dust optical depth are shown in Extended Data Figures 8.

    The expanding shell dominates the dust emission in the region in between the two large circles in Extended Data Figure 8 (see also, the Herschel panel in figure 1). The dust emission in the Huijgens region, directly surrounding the Trapezium stars, is dominated by the dense photodissociation region that separates the ionized gas from the molecular core. As this is not part of the expanding shell, we have excluded this region from our analysis (the region inside the small circle in Extended Data Figure 8). Using theoretical extinction coefficients[46], we obtain for the mass of the shell, M=1700 $M_\odot$. There is considerable uncertainty in the SED fit, associated with the exact choice of the grain emissivity index, $\beta$. Choosing $\beta$ = 1.5 as suggested by the Planck survey[47] decreases $\tau_{160}$ by about 50%, resulting in a derived mass of 900 $M_\odot$, using the same theoretical conversion factor as before. Extended Data Figure 8 shows SED fits towards single points scattered throughout the Orion Nebula. As these are well fitted by $\beta$ = 2, the mass estimate of 1700 $M_\odot$ is appropriate. Finally, we have to make a geometric correction as we have only included the mass in the limb brightened shell. Taking the thickness of the shell as 40% of the radius, we arrive at 2600 $M_\odot$ for the shell mass. Decreasing the thickness of the shell to 20%, the inferred mass increases to 3400 $M_\odot$. We note that, while there is [CII] emission (almost) everywhere, the Veil shows variations in thickness (see below) and the "front" surface seems to be thinner than the limb brightened edge.

    The shell mass can also be estimated from the [$^{13}$CII] lines which are shifted from the [$^{12}$CII] line due to hyperfine splitting. Comparison of the strength of the [$^{13}$CII] hyperfine lines with the (main-component) [$^{12}$CII] line provides the [CII] optical depth and the excitation temperature with an adopted $^{12}$C/$^{13}$C abundance ratio. These then yield the C$^+$ column density and, assuming a C abundance, the H-column density and gas mass in the region can be derived. However, the [$^{13}$CII] line is too weak to be detected in the individual spectra. Only by averaging over a large area are we able to detect the strongest of the three [$^{13}$CII] components, the F = 2-1 [$^{13}$CII], with a relative strength of 0.625 of the total [$^{13}$CII] intensity[48]. This line is offset by 11.2km/s from the [$^{12}$CII] line[48]. The other two [$^{13}$CII] components, F = 1-1 and F = 1-0, lie at 63.2km/s and -65.2km/s, respectively, and are too weak to be detected in our data. Extended Data Figure 9 shows the spectrum averaged over the shell (the area indicated by the two large circles in Extended Data figure 8). Again, we have excluded the emission from the dense PDR directly behind the Trapezium (the small circle in Extended Data figure 8). We also show the spectrum averaged over the brighter South-Eastern portion of the shell. Both these two averages reveal the

presence of the [$^{13}$CII] hyperfine component. The integrated intensity of the [$^{i}$CII] line is given by[49]

$$I_i = B(T_{ex})\left(\frac{\delta v_D}{c}\right)vf(\tau_i)$$

where $i$ indicates the isotope, $T_{ex}$ is the excitation temperature, $\delta v_D$ the Doppler width, $c$ the speed of light and $v$ the frequency of the line. The factor, $f(\tau_i)$, takes optical depth effects into account. This factor is given by

$$f(\tau_i) = 0.428(E_1(2.34\tau_i) + ln[2.34\tau_i] + 0.57721)$$

where $E_1$ is the exponential function. The [$^{13}$CII] line is optically thin and, in that case, $f(\tau_{13})$ is well approximated by $0.625\tau_{13}$, where the numerical factor takes the hyperfine strength into account. The [$^{12}$CII] line is optically thick and the logarithmic term will dominate. We can relate the optical depth of the two isotopes through the carbon isotopic abundance measured for Orion ($^{12}$C/$^{13}$C=67[50]) and wind up with two equations in two unknowns.

We will analyze here the observed intensity ratio of the two lines average over the bright portion of the shell as that is best determined. This results in $\tau_{12} = 3.5 \pm 1.0$ and an excitation temperature of 44 K. With the observed line width (5km/s), we arrive then at $^{12}$C$^+$ column density of $3.5\times10^{18}$ cm$^{-2}$. Assuming that all the gas phase C is ionized – appropriate for a PDR – and a gas phase carbon abundance[51] of $1.6\times10^{-4}$, and extrapolating these values to the full limb-brightened shell, the inferred hydrogen mass is 1700±400 M$_\odot$. Given the large uncertainties involved, this is gratifyingly similar to the estimate from the dust emission. Correcting for geometry, adopting a shell thickness of 40% of the radius, results in a total mass of 2600 M$_\odot$. This is an upper limit on the total mass as the brightest portion of the [CII] shell might be characterized by a larger than average column density. Also, while the derived optical depth is not sensitive to beam dilution, the derived excitation temperature is sensitive to this. If we were to adopt an excitation temperature of 100K, the inferred mass, corrected for geometric effects, would only be 1700 M$_\odot$.

The mass of the shell can also be estimated from the observed optical extinction towards the ionized gas in the M42 HII region as this extinction is dominated by the gas in the Veil[52]. From a comparison of the MUSE Hβ map with the radio emission, an average visual extinction of 1.8 mag is derived[53]. Using the theoretical extinction law[46] for $R_V$=5.5, this corresponds to an average H column density of $2\times10^{21}$ cm$^{-2}$. This is a factor of 2 less than the H column density directly measured towards θ$^1$ Ori C ($N_H$ =$4.8\times10^{21}$ cm$^{-2}$)[11]. Adopting this latter value, we arrive at a total mass for the shell ($2\pi R^2 N_H \mu m_H$ with $m_H$ the mass of hydrogen and $\mu$ the mean molecular weight) of 1300 M$_\odot$. The observed extinction is very patchy and the optical nebula seems to be located behind a relative thin portion of the Veil[52]. Hence, we consider this a lower limit. As we consider the dust estimate the most direct measurement of the shell mass, we will adopt 2600 M$_\odot$ for the mass of the Veil in our analysis, recognizing that this value is uncertain by a factor 2.

**Data availability Statement:** The data sets analyzed during the current study are available through the SOFIA data archive at https://dcs.sofia.usra.edu/dataRetrieval/SearchScienceArchiveInfo.jsp and can be retrieved by searching for the PI (Alexander Tielens) and instrument (GREAT).

**Extended Data**

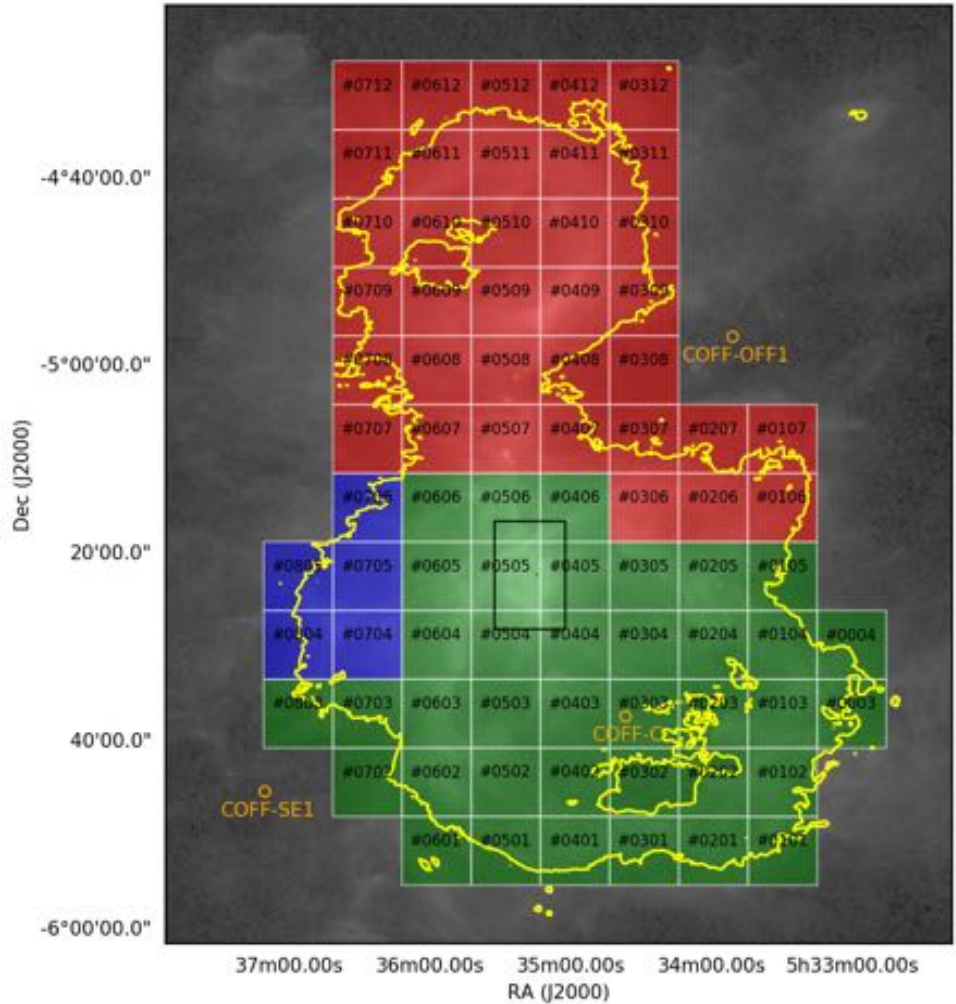

**Extended Data Figure 1: Outline of the region mapped in the [CII] 1.9 THz line with upGREAT on SOFIA.** The 78 tiles indicated were used to construct the final map. Background image is a 70 micron Herschel/PACS dust emission. The yellow contours correspond to an approximated far-UV radiation field of $G_0$=50 (in Habing units). The color of each tile indicate its corresponding OFF position, blue tiles use the COFF-SE1 position, red tiles COFF-OFF1 and green tiles COFF-C. Each square tile has a side length of 435.6 arcseconds. The black box at the center indicates the region mapped by the single pixel Herschel/HIFI instrument in a time of 9 hours[54]. The total observing time for the SOFIA/upGREAT map was 42 hours.

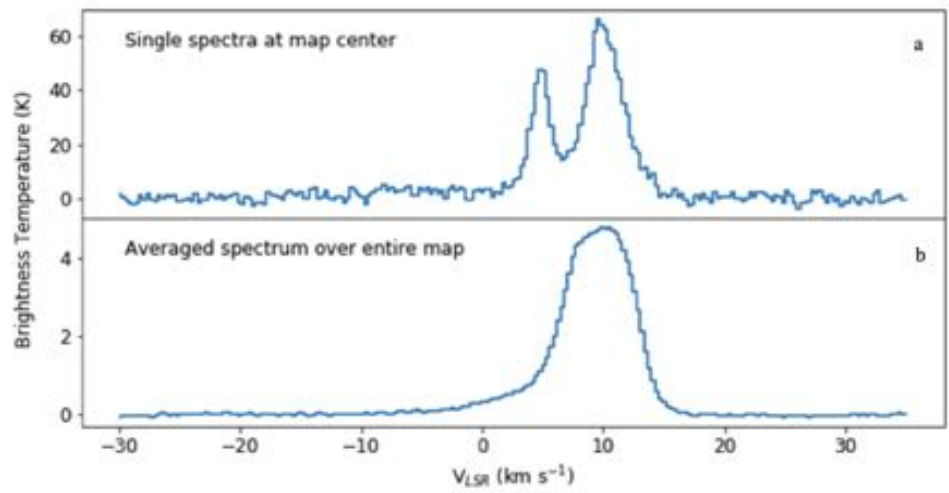

**Extended Data Figure 2: Sample [CII] 1.9THz spectra in our data cube. a)** Spectrum obtained at the map center (RA: 5h35m17s Dec:-5dm22d16.9). **b)** Averaged spectrum over the entire map.

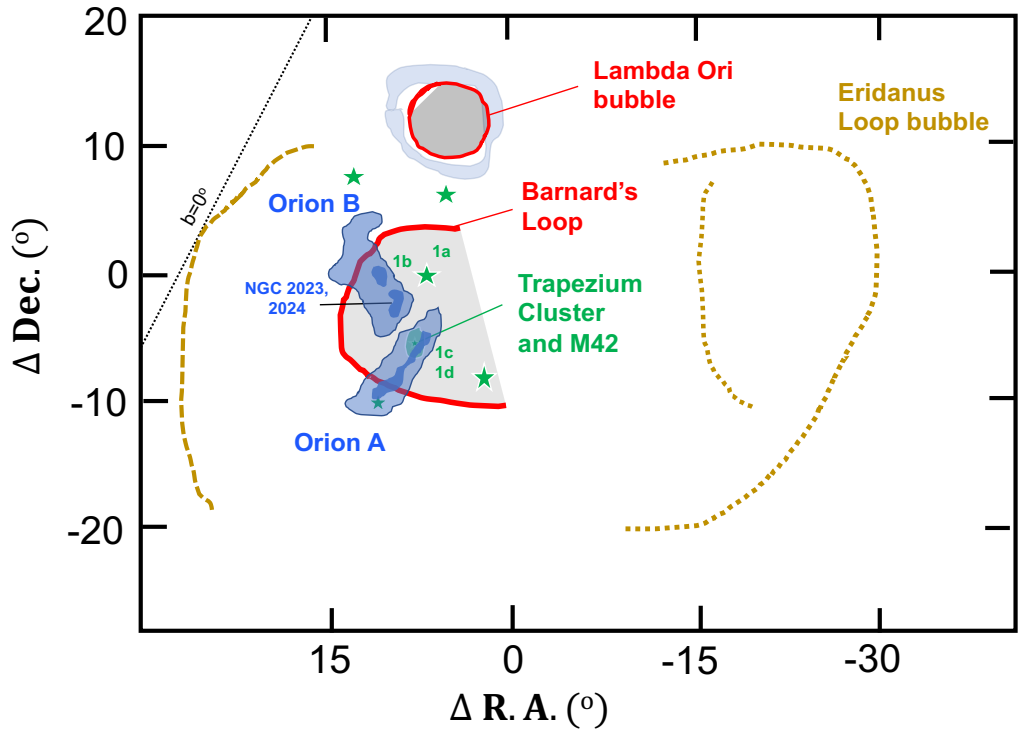

**Extended Data Figure 3**: **Schematic overview of the large scale (~350 pc) structure of Orion.** We mark with green stars the locations of the massive stars spanning up the Orion constellation (shoulders and knees, but the belt is indicated by a single star; M42 is at the tip of the sword), in blue outline the two giant molecular clouds A & B, in green filled trace the prominent HII regions including M42 powered by the trapezium cluster, in red Barnard's loop, which is very prominent in Hα. The bubble surrounding λ Ori is also indicated (red=ionized gas, blue swept up molecular shell, and the boundaries of the superbubble (yellow). The locations of the Orion OB subassociations are marked in green.

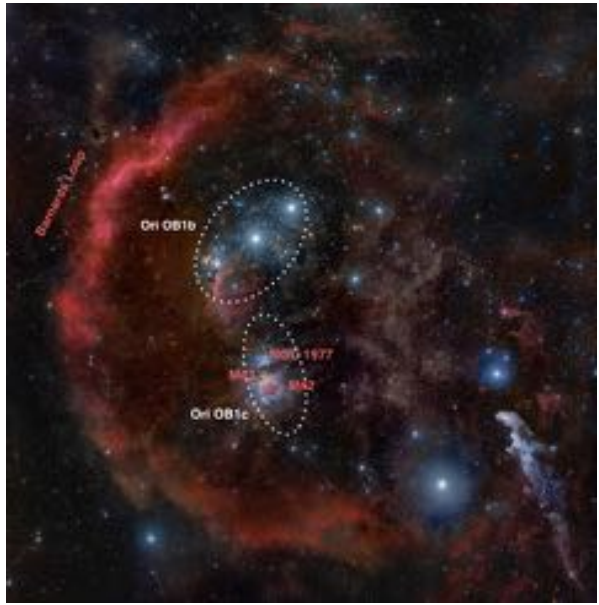

**Extended Data Figure 4: Overview of the star forming region in Orion.** The approximate boundaries of the Orion OB associations Ib, & Ic are indicated by dashed ellipses. The Orion Id association is directly associated with the molecular cloud behind the Orion Nebula, M42. The reddish glow is due to the Hα line originating from recombinations in the ionized gas of Barnard's loop. The belt stars and the knees are obvious. The size of the image is approximately 10 degrees on the sky.

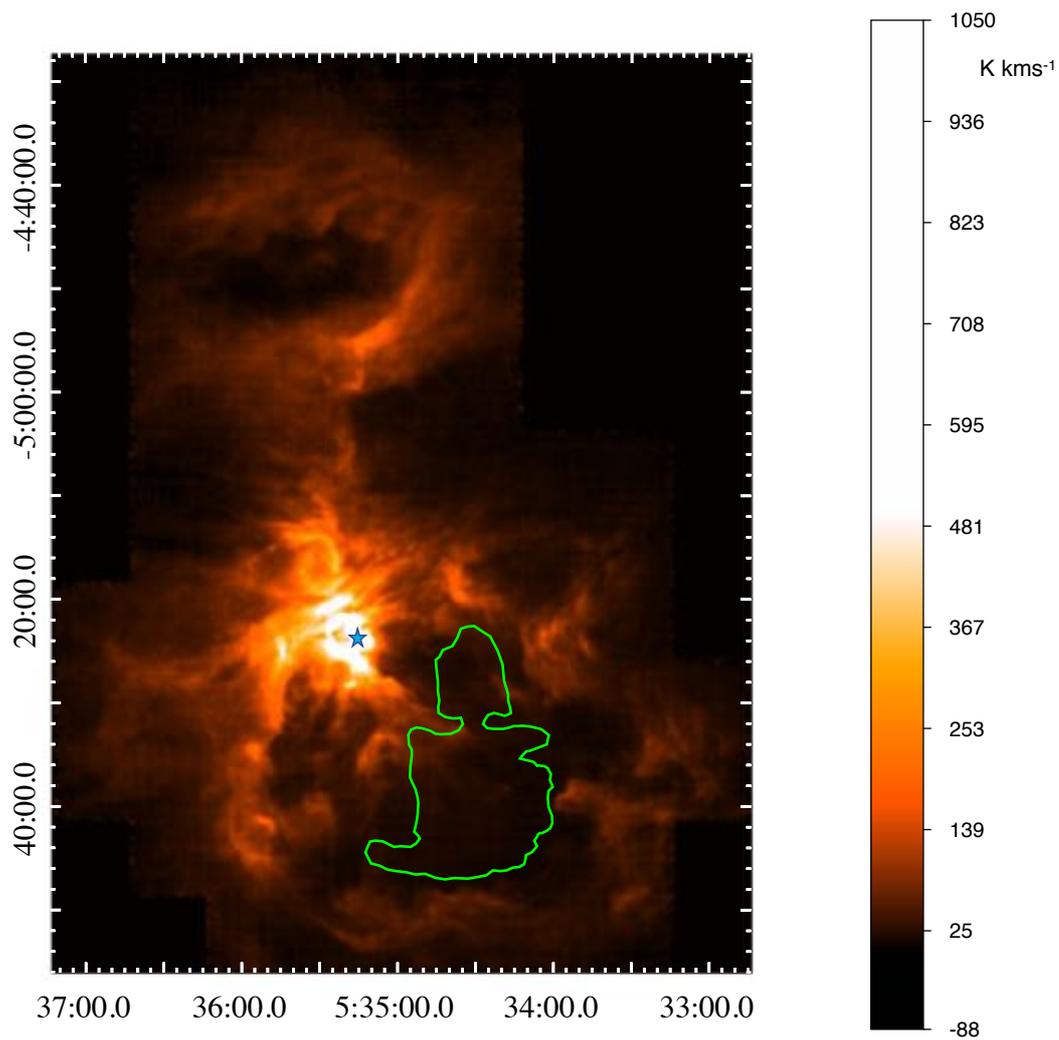

**Extended Data Figure 5: Composite infrared and X-ray views of the Orion region of massive star formation.** The [CII] integrated intensity map is shown on a color scale. The X-ray emission is outlined by a green contour. Likely, the hot gas fills the bubble on its entirety but absorption by the Veil preferentially extinguishes the left side. The position of the star, $\theta^1$ Ori C (RA(2000)=05 35 16.46 Dec(2000)=-05 23 22.8) is indicated by a blue star. Credit: X-ray data XMM-Newton.

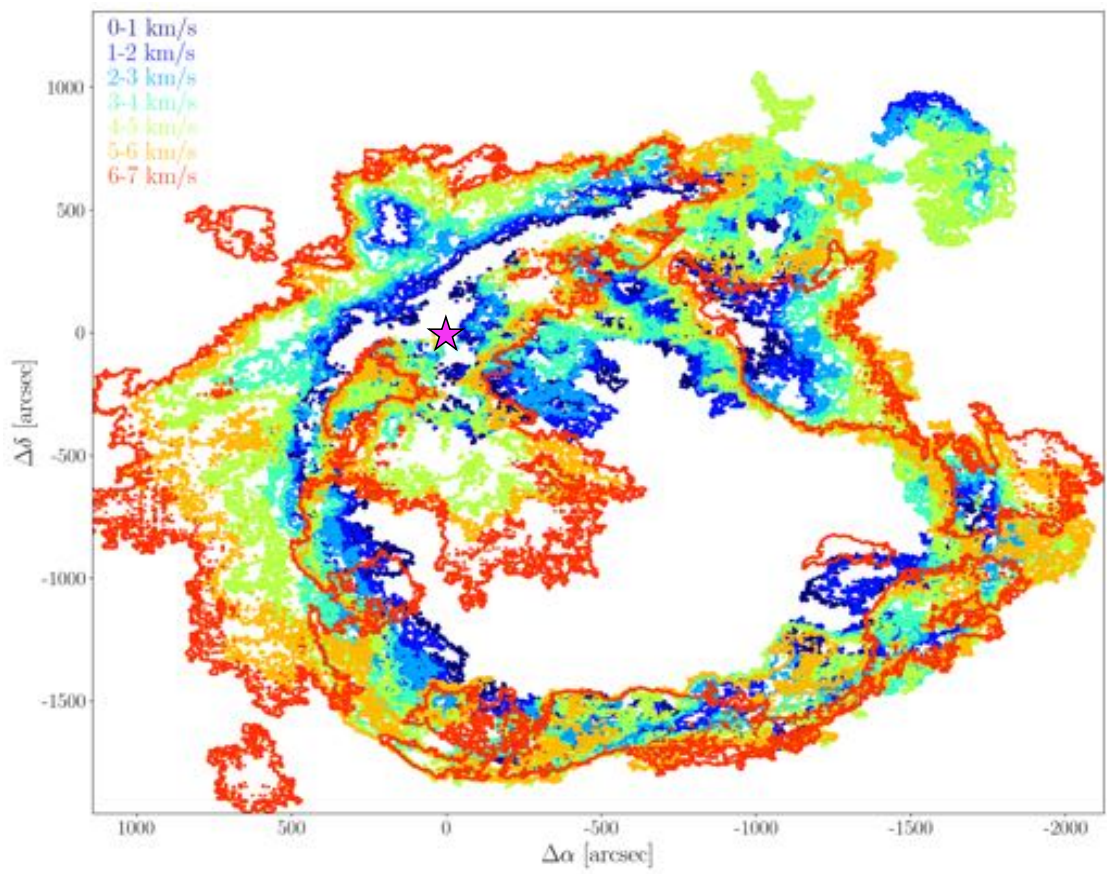

**Extended Data Figure 6: Composite figure showing the [CII] emission in different velocity channels.** Note that with increasing $v_{LSR}$, the shell is displaced outward, away from the bubble center. This is the kinematic signature of an expanding half shell. Each color outlines the emission boundaries of 1 km/s wide channels from $v_{LSR}$ = 0 to 7 km/s. The 0,0 position corresponds to the position of the exciting star, $\theta^1$ Ori C (RA(2000)=05 35 16.46 Dec(2000)=-05 23 22.8) and is marked by a magenta star. Note that in the velocity range, 4 to 7 km/s, [CII] emission associated with the OMC-4 core starts to fill in the bubble interior. OMC-4 is a starforming core near the front of the background molecular cloud and is not part of the Veil bubble.

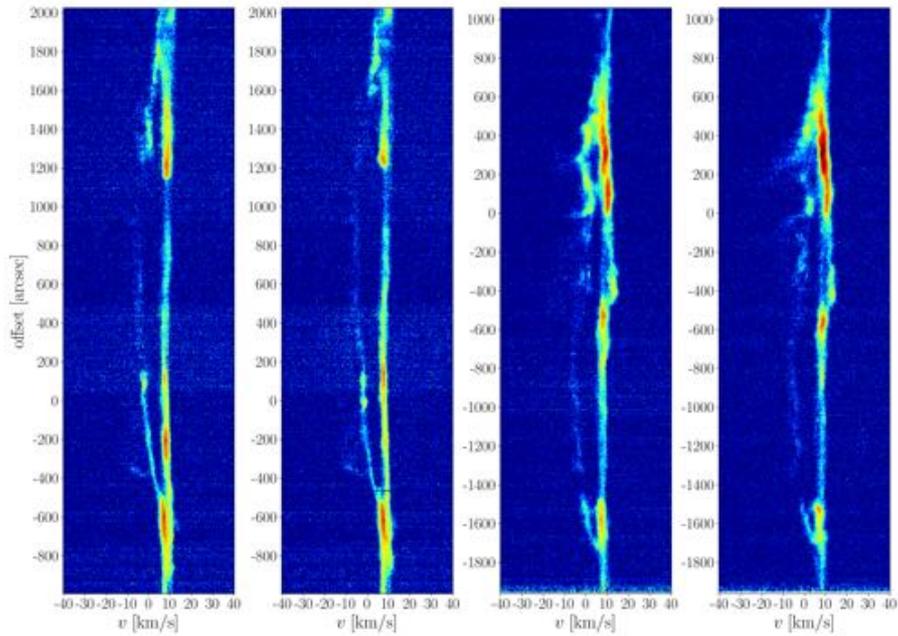

**Extended Data Figure 7: Four exemplary position–velocity diagrams (pv) of the [CII] emission along selected cuts across the Veil.** Each pv diagram exhibits a clear arc structure extending over ~2500", corresponding to the expanding Veil shell (Pabst et al 2018, in preparation). The left two panels are cuts along the horizontal axis. The right two panels show cuts along the vertical axis.

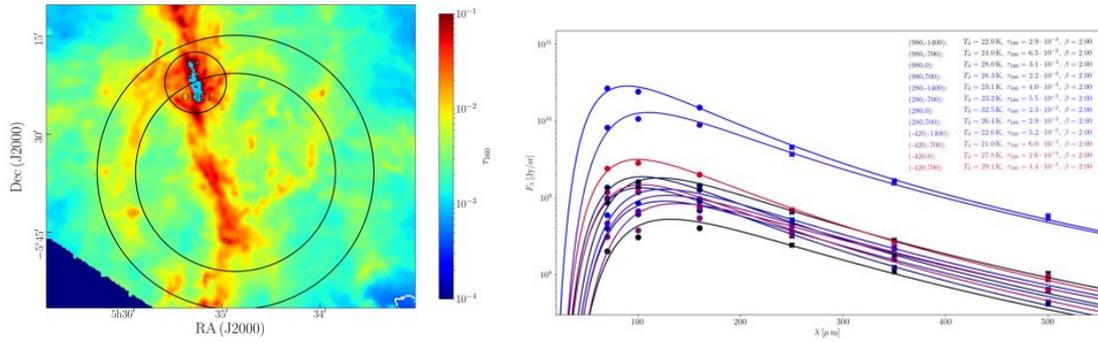

**Extended Data Figure 8: Far infrared dust emission in Orion**. **a)** Optical depth map of the dust emission at 160 μm, $\tau_{160}$, tracing the mass of the shell. The two big circles indicate the extent of the shell used to determine the mass of the limb brightened shell. The small circle, inscribed `OMC1' circumscribes the Huygens region associated with the Trapezium stars. We have estimated the mass that is enclosed between these circles, excluding the OMC1/Huijgens region. **b)** Spectral energy distribution (SED) of the dust emission observed for different positions in Orion. These SEDs are analyzed to determine the dust and gas mass. Data points and curves represent observed SEDs and model fits for β = 2. The legend provides resulting dust temperature $T_d$ and dust optical depth, $\tau_{160}$, values. These SED fits have been analyzed for each spatial point and the resulting optical depth values have been used to construct the map shown in panel a.

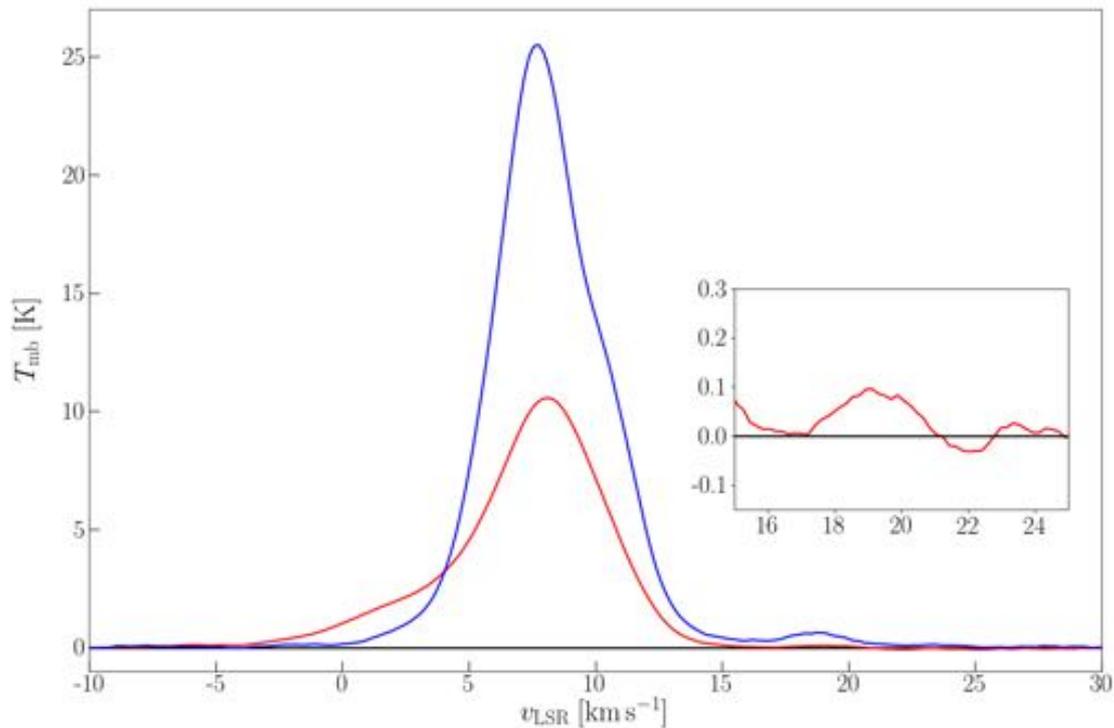

**Extended Data Figure 9: Average spectra from the shell.** These spectra are dominated by the [CII] line from the main isotope and show the weak hyperfine component of $^{13}C^+$ near $v_{LSR} \sim 20$ km/s. This line can be used to estimate the optical depth of the main isotope line and then the mass of emitting gas. The red spectrum corresponds to the area between the two large circles in Fig. 1 but excluding region in the circle encompassing the OMC1/Huygens region. The inset of the spectrum shows the zoom in on the (faint) [$^{13}$CII] line in the average shell spectrum. The blue spectrum is an average over the bright parts in the eastern shell, in the declination range $\delta = -5°35'$ to $-5°45'$.

**Extended Data Table 1: Masses, energetics and luminosities in Orion**

| Component | Mass $M_\odot$ | Energy Thermal $10^{46}$ erg | Energy Kinetic $10^{46}$ erg | Luminosity $L_\odot$ | Ref |
|---|---|---|---|---|---|
| OMC-1 molecular gas | 3000 | 0.6 | 20 | | 17 |
| Veil | 2600 | 3 | 400 | | a |
| Stellar cluster | 1800 | – | | | 18 |
| Ionized gas | 20 | 3 | 6 | | 16,19 |
| Huygens region | 2 | 0.3 | 2 | | 16,19 |
| Hot gas | 0.07 | 10 | – | | 13 |
| Stellar wind | – | – | 500[b] | 200[c] | 14,15 |
| $\theta^1$ Ori C | – | – | – | $2.5 \times 10^5$ | 14 |
| Far-IR dust emission | – | – | – | $6 \times 10^4$ | a |
| [CII] 1.9 THz | – | – | – | 200 | a |
| X-ray | – | – | – | $1.4 \times 10^{-2}$ | 13 |

Notes:
a: This study.
b: Over the calculated lifetime of the bubble.
c: Mechanical luminosity of the wind.